\documentclass[epj,final,nopacs]{svjour}
\usepackage{slashed} 
\usepackage{nccmath} 
\usepackage{times} 
\usepackage{amsmath} 
\usepackage{amssymb} 
\usepackage{verbatim} 
\usepackage{graphicx} 
\usepackage{color} 
\usepackage{booktabs} 
\usepackage{subfigure} 
\usepackage{empheq}
\usepackage{relsize}
 
\newcommand{\nths}{\negthickspace\negthickspace} 

\newcommand{\lp}{\left(} \newcommand{\rp}{\right)} 
\newcommand{\lb}{\left\{} \newcommand{\rb}{\right\}} 
\newcommand{\ls}{\left[}  \newcommand{\rs}{\right]}
\newcommand{\lv}{\left|}  \newcommand{\rv}{\right|}
  
\newcommand{\cd}{\!\cdot\!}
\newcommand{\st}[1]{\slashed{#1}}
\mathchardef\mhy="2D   
\DeclareMathOperator{\Ai}{Ai}

\begin{document}
\title{Furry picture transition rates in the intense fields at a lepton collider interaction point}
\author{A. Hartin}
\institute{II. Institute for Theoretical Physics, University of Hamburg, Luruper Chaussee 149, 22761 Hamburg, Germany}
\mail{anthony.hartin@desy.de}
\date{}
\journalname{}
\abstract{The effect on particle physics processes by intense electromagnetic fields in the charge bunch collisions at future lepton colliders is considered. Since the charge bunch fields are tied to massive sources (the $e^{+}e^{-}$ charges), a reference frame is chosen in which the fields appear to be co-propagating. Solutions of the Dirac equation minimally coupled to the electromagnetic fields reasonably associated with two intense overlapping charge bunches are obtained and found to be a Volkov solution with respect to a null 4-vector whose 3-vector part lies in the common propagation direction. These solutions are used within the Furry interaction picture to calculate the beamstrahlung transition rate for electron radiation due to interaction with the electromagnetic fields of two colliding charge bunches. New analytic expressions are obtained and compared numerically with the beamstrahlung in the electromagnetic field of one charge bunch. The techniques developed will be applied to other collider physics processes in due course. 
} 
\maketitle

\section{Introduction}\label{sect:Intro}
Physics experiments in particle colliders involve the overlap of dense bunches of charged particles which interact to produce final states of interest. The treatment of the individual interactions which produce new particles is of course well known and thoroughly carried out in physics analysis. By comparison, the treatment of collective effects from the strong electromagnetic fields has been limited to beamstrahlung and background pair production processes only. Nevertheless, the overwhelming bulk of physics processes take place in strong overlapping charge bunch fields. There is a lack of a general analysis in this respect and it is the goal of this paper to lay the groundwork of just such a general analysis.\\

At linear colliders, beamstrahlung and coherent pair production have been considered to take place in a constant crossed electromagnetic field. Theoretically, the quasi-classical approximation (QCA) has been used to determine the transition rate \cite{Baier68,Baier69,YokChe91}. A method which does not rely on kinematic approximations and which takes into account the external field non-perturbatively, the Furry picture \cite{Furry51}, was used to calculate the same process transition rates \cite{Ritus72}. The two methods produce identical results after the kinematic approximations of the QCA are accounted for \cite{Hartin09}. \\

The equivalent photon approximation (EPA) has often been utilised to calculate the effect of charged particle fields on collider processes. The main processes producing $e^{+}e^{-}$ pairs are the Landau-Lifshitz and Bethe-Heitler processes. At linear colliders these processes, along with the Breit-Wheeler process make up the incoherent pair production \cite{YokChe91}. The applicability of the EPA depends on the size of the impact parameter compared to the Compton wavelength and is not suitable for multiphoton radiation \cite{AlsBau97}. The inapplicability of the EPA also includes beamstrahlung, neutrino and heavy lepton pair production \cite{Budnev75}. \\

In this paper, the effect of charge bunch fields on general processes of any order, without kinematic approximations, will be considered. The natural theoretical framework will be the Furry picture which requires solutions of the field equations embedded in the electromagnetic fields of the charge bunches. There appears to be, up to the present, no consideration of the case of overlapping, counter-propagating charged bunches. This is the particular problem to be addressed in this paper. \\

Exact solutions of the fermion equations of motion with external electromagnetic field are known for plane wave fields, centrally symmetric fields and a few cases of a combination of fields \cite{BagGit90}. Exact solutions for spin 0 and spin 1 charged particles embedded in plane wave fields are also known \cite{Kurilin99}. \\

For the case of two plane waves, exact fermion solutions have previously been found for the restricted case of collinear fields with orthogonal vector potentials \cite{Lyulka75,Pardy06}. Volkov solutions of combinations of laser fields of different modes have been used in the study of photon emission and pair photoproduction \cite{NarFof96}. The case of two anti-collinear, circularly polarised fields was considered by \cite{BerVal73} for the Klein-Gordon equation, and by \cite{SenGupta67} for the Dirac equation. \\

The case of overlapping charge bunch fields dealt with in this paper, will be considered in a reference frame in which the required computations are simplest. The Lorentz transformation of photons reveals that two, free electromagnetic fields either co-propagate in all reference frames or they can be transformed to a frame in which they anti-propagate. The fields associated with charge bunches, however, are tied to massive sources. It is always possible therefore to adopt a reference frame in which the charge bunches and their associated fields appear to co-propagate. \\

In section \ref{sect:collbunch} it will be argued that it is not possible to assume that the vector potentials of the two charge bunch fields are orthogonal. Lepton colliders bring together bunches of opposite charge whose 3-potentials are generally anti-parallel. However, the bunch collisions are not perfectly head-on, and the bunches distort during the collision. The 3-potentials of the two bunches, though co-planar, must otherwise be allowed to have general orientation in analytic calculations. \\

The solutions sought in section \ref{sect:diracsol} will be those of charged spin 1/2 particles in two co-propagating plane wave fields of otherwise general form. In such a combination of fields, a Volkov solution with respect to a null 4-vector whose unit 3-vector part lies in the common propagation direction, will be employed. In order to be certain this procedure is valid, an explicit proof is given in \ref{sect:App1}. \\

In order to obtain commutation relations and quantisation of wavefunctions for use in the calculation of transition rates, orthogonality and completeness of the exact fermion solutions in two fields is necessary. The proof can be considered to be the same as that for the Volkov solution for a fermion in a single plane wave field. It can then be assumed that the commutation relations and quantisation within the Furry picture are canonical. \\

The explicit fermion solutions in two charge bunch fields will be applied directly to the calculation of the beamstrahlung transition rate in section \ref{sect:beam}. A new analytic form for this transition rate will be obtained and compared to the usual beamstrahlung expression. \\

This paper intends to lay the groundwork for a consideration of general collider phenomenology in the overlapping electromagnetic fields of colliding charge bunches. Given the strength of the fields expected at future linear colliders, this analytic work is considered essential, particularly for precision calculations. The next steps in this analytic program, following on from this paper, will be discussed in section \ref{sect:concl}. \\

This paper works in natural units and the electromagnetic gauge is fixed by a combination of the Lorenz gauge and setting the time component of the 4-potential to zero,

\begin{gather}\label{eq:notn}
c=\hbar=4\pi\epsilon_0=1, \quad e=\sqrt{\alpha} \notag\\
\partial^\mu A^e_\mu=0, \quad A^e\equiv (0,\vec{A}^e)
\label{eqgauge}\end{gather}

Superscripts, $\psi^\pm,u^\pm$ will refer to the positive and negative energy solutions of the Dirac equation. Subscripts consisting of Greek letters will refer to 4-vector components in Cartesian coordinates and $\pm$ subscripts refer to linear combinations of 4-vectors and functions. Two or three component vectors will be represented by, $\vec{x}$.

\section{The external field of two colliding charge bunches}\label{sect:collbunch}


Lepton colliders, from the point of view of the laboratory reference frame, bring together two ultra relativistic, Gaussian charge bunches with a flat transverse profile. Since it is the aim of this paper to consider particle processes in the overlapping fields of both charge bunches, it is sensible to examine the collision in a reference frame that simplifies the calculation. \\

\begin{figure}[h] 
\centerline{\includegraphics[width=0.45\textwidth]{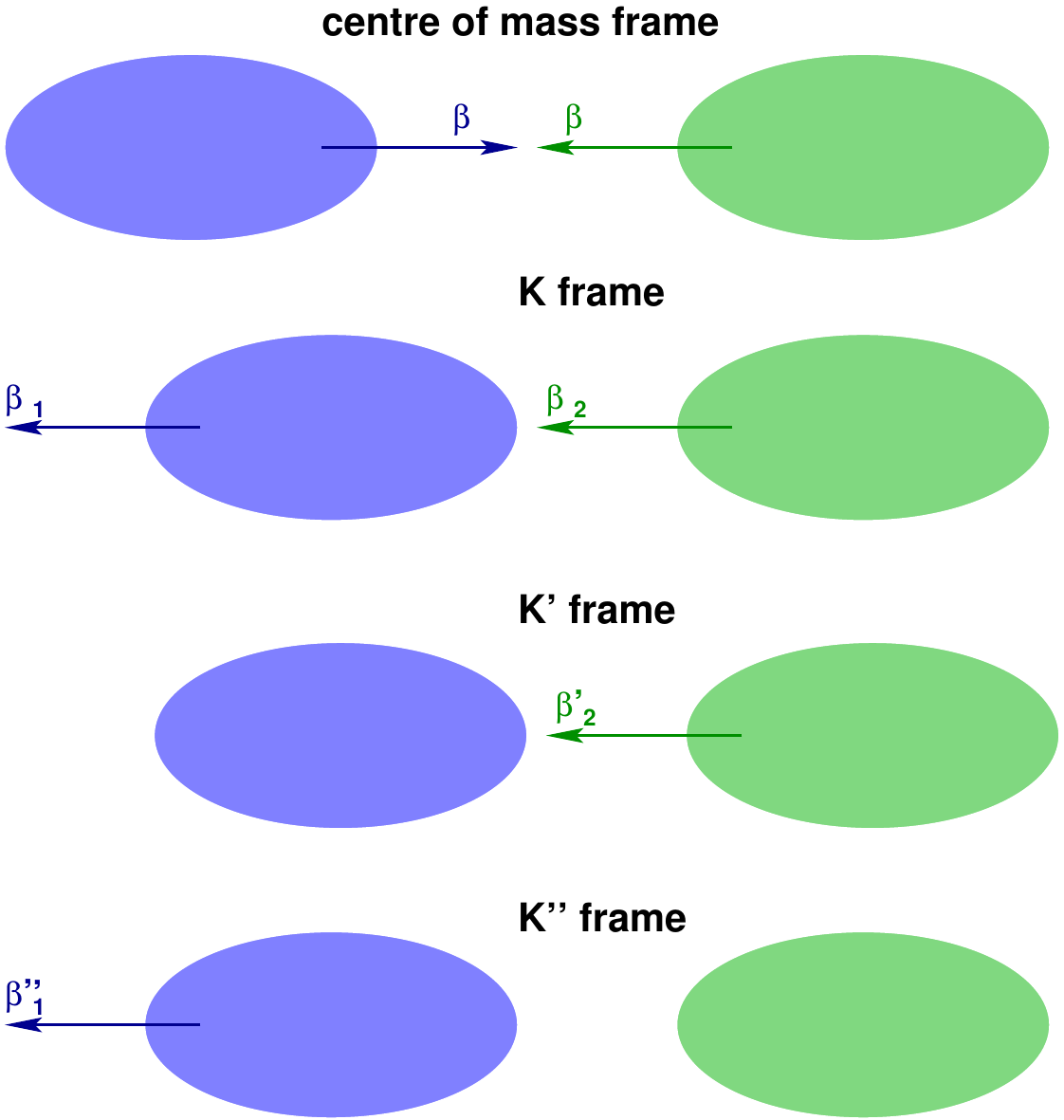}}
\caption{\bf Apparent charge bunch motion for an observer in center of mass, K, K' and K'' reference frames.}\label{fig:refframes}
\end{figure}

By moving the point of view relativistically, it is possible to adopt reference frames in which either of the bunches appear to be at rest (the $K'$ and $K''$ frames). By moving the observer further the $K$ frame can be adopted in which both bunches appear to be moving relativistically and collinearly (figure \ref{fig:refframes}). \\

\begin{figure}[h] 
\centerline{\includegraphics[width=0.5\textwidth]{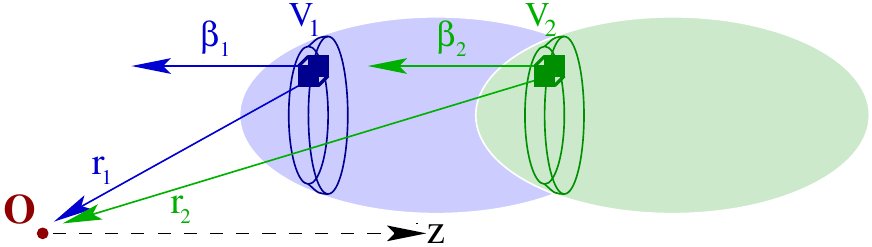}}
\caption{\bf Volume elements and associated vectors in the K frame.}\label{fig:voleles}
\end{figure} 

In order to study particle physics processes in the fields of the bunches it is desirable to know the nature of the fields seen at an observer point $O$ in the in the K frame. Infinitesimally small volume elements,$V_1$ and $V_2$, containing charge $q_1$ and $q_2$ move with relativistic beta $\vec{\beta}_1,\vec{\beta}_2$ collinearly along the z-axis (figure \ref{fig:voleles}). The vectors $\vec{r}'_1,\vec{r}''_2$, in the $K'$ and $K''$ frames respectively, transform to transverse ($\vec{r}_{1\perp},\vec{r}_{2\perp}$) and longitudinal ($\hat{\vec{z}}$) vectors in the $K$ frame as,

\begin{gather}\label{eq:rtransform}
\vec{r}'_1=\vec{r}_{1\perp}+\gamma_1(z-v_1t)\hat{\vec{z}},\quad \vec{r}''_2=\vec{r}_{2\perp}+\gamma_2(z-v_2t)\hat{\vec{z}}
\end{gather}

so that the electric fields in each element $V_1$ and $V_2$, within the $K'$ and $K''$ frames respectively, are \\

\begin{gather}\label{eq:K'K''frameE}
\vec{E}'_1=\mfrac{q_1\vec{r}'_1}{|\vec{r}'_1|^3}=\mfrac{q_1\vec{r}_{1\perp}+\gamma_1(z-v_1t)\hat{\vec{z}}}{\ls|\vec{r}_{1\perp}|^2+\gamma_1^2(z-v_1t)^2\rs^{3/2}} \notag\\[4pt]
\vec{E}''_2=\mfrac{q_1\vec{r}''_2}{|\vec{r}''_2|^3}=\mfrac{q_2\vec{r}_{2\perp}+\gamma_2(z-v_2t)\hat{\vec{z}}}{\ls |\vec{r}_{2\perp}|^2+\gamma_2^2(z-v_2t)^2\rs^{3/2}}  
\end{gather}

The transformation of the fields themselves to the $K$ frame is achieved with the relations (\cite{Jackson75}, section 11.10)

\begin{gather}\label{eq:KframeE1E2}
\vec{E}_1=\gamma_1\vec{E}'_1-\mfrac{\gamma_1^2}{1+\gamma_1}\vec{\beta}_1(\vec{\beta}_1\cdot\vec{E}'_1);\quad \vec{B}_1=\vec{\beta}_1\times\vec{E}_1 \notag\\
\vec{E}_2=\gamma_2\vec{E}''_2-\mfrac{\gamma_2^2}{1+\gamma_2}\vec{\beta}_2(\vec{\beta}_2\cdot\vec{E}''_2);\quad \vec{B}_2=\vec{\beta}_2\times\vec{E}_2
\end{gather}

The combined electric and magnetic fields of the two volume elements seen by the observer in the $K$ frame is,

\begin{align}\label{eq:KframeE}
\vec{E}&=\vec{E}_1+\vec{E}_2 \notag\\
&=\mfrac{\gamma_1(q_1\vec{r}_{1\perp}+(z-v_1t)\hat{\vec{z}})}{\ls|\vec{r}_{1\perp}|^2+\gamma_1^2(z-v_1t)^2\rs^{3/2}}+\mfrac{\gamma_2(q_2\vec{r}_{2\perp}+(z-v_2t)\hat{\vec{z}})}{\ls|\vec{r}_{2\perp}|^2+\gamma_2^2(z-v_2t)^2\rs^{3/2}} \notag\\[4pt]
\vec{B}&=\vec{\beta}_1\times\vec{E}_1+\vec{\beta}_2\times\vec{E}_2
\end{align}

For ultra relativistic bunches, $\beta_{1,2}\rightarrow 1,\; \gamma_{1,2}\gg 1$. The electric and magnetic fields take on the form of sharply peaked transverse pulses centered around the z position of the volume elements. The magnetic fields become of the same magnitude as, and perpendicular to, the electric fields. \\

Physics processes occur as the bunches collide and volume elements $V_1$ and $V_2$ overlap. The true fields that exist during the bunch collision are complicated, as the bunches electromagnetic interactions produce field distortions. Analytic calculations of transition rates that take into account these true fields exactly, are unfeasible. The calculation strategy is to determine transition rates of physics processes of interest within each volume element. To then calculate the transition rate across the whole bunch collision, a simple sum is required for all the volume elements encompassing the bunch collision. \\

This strategy allows the electromagnetic fields within the volume elements to be approximated. If the volume elements are small and relativistic enough, the field's longitudinal components can be neglected and the field strengths can be considered constant. Such approximations allow the electromagnetic fields of each volume element to be described by a constant crossed field.  The 4-potential representing two such overlapping volume elements is a sum of two terms. Each term contains information about the propagation direction of each bunch, which can vary independently depending on the chosen reference frame

\begin{gather}\label{eq:ccfield}
A^e_\mu=ea_{1\mu}(k_1\cd x)+ea_{2\mu}(k_2\cd x) \\
a_{1\mu}=(0,\vec{a}_1),\; a_{2\mu}=(0,\vec{a}_2) \notag \\
k_{1\mu}=(\omega_1,\vec{k}_1),\; k_{2\mu}=(\omega_2,\vec{k}_2) \notag 
\end{gather}

The bunch collision may not necessarily be centred with respect to the centre of each bunches transverse cross-section. So, the three potential of the field of each volume element may have any direction as long as its transverse to the propagation direction of its own bunch. The 4-vectors $k_1,k_2$ are null 4-vectors whose 3-vector lies along each bunch's propagation direction. \\

In the calculation of Lorentz invariant transition rates of processes embedded in the fields, scalar products will be formed from the electromagnetic field tensor. These typically appear as a dimensionless parameter,

\begin{gather}\label{eq:Upsilon}
\Upsilon\equiv\sqrt{(eF_{\mu\nu}p_\nu)^2}/m^3
\end{gather}

For the constant crossed fields considered in this paper (equation \ref{eq:ccfield}), the invariant quantitiy $\Upsilon$ is,

\begin{gather}\label{eq:Upsilon2f}
\Upsilon \equiv e\mfrac{\sqrt{(a_{1\mu}\;k_1\cdot p+a_{2\mu}\;k_2\cdot p})^2}{m^3}
\end{gather}
 
In the interest of precision, and since the electromagnetic fields of highly compressed charge bunches are intense, it is desirable to adopt a QFT which takes into account the charge bunch fields exactly. The most useful QFT to use is the Furry picture which is non-perturbative with respect to the external 4-potential. The Furry picture will be used to generate equations of motion for spin 1/2 particles. \\

The solutions obtained from the equations of motion will then be applied to the particular case of two constant crossed electromagnetic fields associated with ultra-relativistic charge bunch collisions. The solutions will be used to calculate a new analytic expression for the beamstrahlung transition rate in both charge bunches, which reduces to the regular expression in appropriate limits.

\section{Dirac equation solutions in collinear fields}\label{sect:diracsol}

In the Furry picture \cite{Furry51,Schweber62,JauRoh76,Moort09}, the electromagnetic gauge field is separated into a part $A_\mu$ that interacts perturbatively with coupled fields, and an external part $A^e_\mu$ whose interaction with coupled fields is calculated exactly. Thus, the Lagrangian density for a charged spinor of mass and charge $m,e$ is \cite{FraGitShv91},

\begin{gather}\label{diraceq}
\mathcal{L}_{\text{DIRAC}}^{\text{FP}}=\bar\psi(i\slashed{\partial}-e\st{A}^e-m)\psi \end{gather}

 from which the Euler-Lagrange equation gives the equation of motion,

\begin{gather}\label{eqofmotion}
\lp i\slashed{\partial}-e\st{A}^e-m \rp \psi_{\text{FP}}=0
\end{gather}

For an external potential described by a single plane wave, the solution of the minimally coupled Dirac equation (equation \ref{eqofmotion}) is the well known Volkov solution \cite{Volkov35}, $\Psi_\text{V}$, which differs from the usual free fermion solutions by the inclusion of an extra phase and extra spinor $E_p$. For an electron of momentum $p_\mu\!=\!(\epsilon,\vec{p})$ in an external field of 4-momentum $k_\mu\!=\!(\omega,\vec{k})$ and with $u_{rp}^\pm$ being the positive and negative energy Dirac spinors with spin r, the positive and negative energy Volkov solutions are,

\begin{gather}\label{eq:Volkov}
 \Psi^{\pm}_\text{V}(\phi)= \sqrt{\small\mfrac{m}{2\epsilon(2\pi)^3}}\,E_p^\pm(\phi)\; u^\pm_{rp}\,e^{\mp i p\cdot x } \\
E_p^\pm(\phi)\equiv\ls 1 \mp \mfrac{e\slashed{A}^e\st{k}}{2(k\cd p)}\rs e^{\mp i\mathlarger{\int}^\phi{\mfrac{2eA^{e}(\xi)\cd p \mp e^2A^e\!(\xi)^2}{2k\cd p}}d\xi}, \; \phi\equiv k\cd x \notag
\end{gather}

The fermion solutions in two co-propagating plane wave fields, $\psi^{\pm}_{\text{OCP}}$, were obtained for the special case of orthogonal 3-potentials \cite{Lyulka75,Pardy06} and applied to various combinations of laser fields. The solution obtained contains a product of Volkov $E_p$ functions,

\begin{gather}\label{eq:psiOCP}
A^e_\mu=A_{1\mu}(\phi_1)+A_{2\mu}(\phi_2), \quad A_1\cd A_2=k_1\cd k_2=0 \notag\\[4pt]
\phi_{1,2}\equiv k_{1,2}\cd x \notag \\[4pt]
\psi^{\pm}_{\text{OCP}}(\phi_1,\phi_2) = \mfrac{E^\pm_{1p}(\phi_1)\; E^\pm_{2p}(\phi_2)\; u^\pm_{rp}}{\sqrt{2\epsilon(2\pi)^3}}\,e^{\mp i p\cdot x} \\
E_{ip}^\pm(\phi_i)\equiv\ls 1 \mp \mfrac{e\slashed{A}_i^e\st{k}}{2(k_i\cd p)}\rs e^{\mp i\mathlarger{\int}^{\phi_i}\mfrac{2eA_i(\xi_i)\cdot p \mp e^2A_i\!(\xi_i)^2}{2k_i\cdot p}d\xi_i} \notag
\end{gather}

However, \cite{Lyulka75,Pardy06} were overly constrictive, since the requirement of orthogonal 3-potentials is not necessary. The 4-potential of two (and indeed any number of) collinear plane-wave fields of 4-momenta $k_{1\mu}$, $k_{2\mu}$ can be rewritten as a function of a null 4-vector $n_\mu$ whose 3-vector part is parallel to the common propagation direction of the collinear fields,

\begin{gather}
k_{1\mu}=\omega_1\; n_\mu,\quad k_{2\mu}=\omega_2\; n_\mu,\quad n_\mu\equiv (1,\vec{n}), \quad \vec{n}^2=1 \notag\\
A_{1\mu}(\phi_1)+A_{2\mu}(\phi_2) \equiv A^{e}_\mu(n\cd x)
\end{gather}

One then posits that the general positive and negative energy solutions $\psi^\pm_{\text{CP}}$ of the Dirac equation minimally coupled to the combination of two co-propagating external fields $A^e_\mu$, are Volkov solutions with respect to the 4-vector $n_\mu$,

\begin{gather}\label{eq:finalpsicp}
\psi^{\pm}_{\text{CP}}(n\cd x) = \sqrt{\mfrac{m}{2\epsilon(2\pi)^3}} J^\pm_{px}\; u^\pm_{rp}\, e^{\mp i\,p\cdot x} \\
J^\pm_{px}\equiv\ls 1 \mp\mfrac{\st{A}^e\!(n\cd x)\st{n}}{n\cd p}\rs\, e^{\mp i\mathlarger\int^{n\cdot x}\small\mfrac{2A^e(\xi)\cd p \mp A^e(\xi)^2}{2n\cd p}d\xi} \notag
\end{gather}

An explicit proof that this is indeed the case for two collinear plane wave fields of general form, is presented in \ref{sect:App1}. \\

To use this solution in the calculation of general particle physics processes in the fields of the two overlapping bunches, canonical commutation relations and quantisation are achieved within the Furry picture in the usual fashion \cite{FraGitShv91}. A prerequisite for the quantisation is the requirement of orthogonality and completeness of the solutions $\psi^\pm_{\text{CP}}$. \\

\cite{Ritus79} proved the orthogonality of the Volkov solutions for fermions in a single plane wave electromagnetic field on the equal time hyperplane. The proof proceeded by finding a new basis for the Volkov spinor which resulted in the Volkov phase when acted on by its hermitian conjugate. \cite{Filip85} achieved the same result by rewriting the standard Volkov spinor in terms of the derivative of the Volkov phase. Completeness of the Volkov solutions on equal time hyperplanes was proven by \cite{BocFlo10}. \\

Also required for the calculation of transition rates, is an explicit form of the general solution for the constant crossed fields expressed in equation \ref{eq:ccfield}. The explicit solutions are

\begin{gather}
\psi^{\pm}_{\text{CP}}(n\cd x) = \sqrt{\mfrac{m}{2\epsilon(2\pi)^3}}\; J^\pm_{px}\; u^\pm_{rp}\, e^{\mp i(p\cdot x)} \\
J^{\pm}_{px} \equiv \ls 1\mp\mfrac{(\omega_1\st{a}_1+\omega_2\st{a}_2)\;\st{n}\; (n\cd x)}{n\cd p}\rs \hspace{2cm} \notag\\
\quad\quad\centerdot \; e^{\mp i\small\mfrac{(\omega_1a_1\cd p+\omega_2a_2\cd p)\,(n\cd x)^2 \mp \frac{1}{3}(\omega_1a_1+\omega_2a_2)^2(n\cd x)^3}{2n\cd p}} \notag
\end{gather}

In the next section, the explicit fermion solutions $\psi^\pm_{\text{CP}}$ will be applied to the beamstrahlung process, which is a 1-vertex processes within the Furry picture. The beamstrahlung transition rate for two bunch fields will be compared with that for one bunch field.

\section{The beamstrahlung in overlapping charge bunch fields}\label{sect:beam}

\begin{figure}[h] 
\centerline{\includegraphics[width=0.25\textwidth]{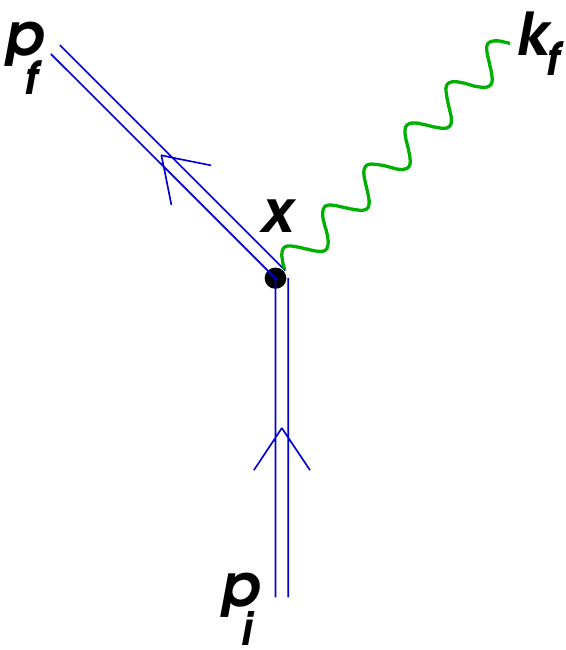}}
\caption{\bf The beamstrahlung Feynman diagram.}\label{fig:beamstr}
\end{figure}

The beamstrahlung (figure \ref{fig:beamstr}) is the photon radiation from individual charged particles at a collider interaction point due to interaction of the charged particles with the fields from the colliding bunches. Usually, the field of the particle's own bunch is neglected and only the interaction with the oncoming bunch field is considered. Here, the beamstrahlung transition rate in both bunch fields will be calculated with the aid of the fermion solutions obtained in section \ref{sect:diracsol}. \\

The matrix element of the beamstrahlung process can be written down with the aid of the Feynman diagram (figure \ref{fig:beamstr}),

\begin{gather}
M_{fi}=-ie\int d^4x\; \bar \psi^{+}_{\text{CP}}(x,p_f)\,\gamma_\mu \psi^{+}_{\text{CP}}(x,p_i)
\end{gather} 

One could proceed, as usual, in writing the square of the matrix element and integrating over space-time and wave packet momentum. However, much of the work has been done for this process, where the transition probability was obtained for any number of overlapping external fields with a common propagation direction \cite{Hartin11a}. Writing the transition propbability as a rate, adapting the result for the 4-vector $n_\mu$, and allowing for a relativistic initial particle expressed by $\gamma\equiv\epsilon_i/m$, the transition rate per unit time is,

\begin{multline}\label{eq:HICSWfinal1st}
W_{\text{beam}}=-\mfrac{e^2}{4\pi\gamma}\int_{-\pi L}^{\pi L} \mfrac{d\phi_w}{2\pi L} \int_0^{\infty} \nths\mfrac{du}{(1+u)^2} \int_{-\infty}^{\infty} \mfrac{i\pi d\lambda}{2\lambda}\\
\centerdot \lb 2+\mfrac{1+(1+u)^2}{2(1+u)} F_1(\lambda\!+\!\phi_w,\phi_w)\rb \\
\centerdot\exp\lp i\mfrac{m^2u}{2(n\cdot p_i)}\lp\lambda-\mfrac{F_2(\lambda+\phi_w,\phi_w)^2}{\lambda}+F_3(\lambda\!+\!\phi_w,\phi_w)\rp \rp
\end{multline}

The functions $F_1,\,F_2,\,F_3$ depend on the external field 4-potential through the relations,

\begin{gather}
F_1(\phi_v,\phi_w)\equiv -\mfrac{e^2}{m^2}\ls A^e(\phi_v)-A^{e}(\phi_w) \rs^2 \notag\\[4pt]
F_2(\phi_v,\phi_w)\equiv \mfrac{e}{m}\lv \int_{\phi_w}^{\phi_v}\nths A^{e}(\phi)d\phi \rv \\
F_3(\phi_v,\phi_w)\equiv -\mfrac{e^2}{m^2}\int_{\phi_w}^{\phi_v}\nths A^e(\phi)^2\,d\phi \notag
\end{gather}

where the 4-potential itself is a sum of the constant crossed fields of two ultra-relativistic charge bunches and is a function of the unit vector scalar product $n.x$ via the relation,

\begin{gather}
A^e_\mu(\phi)=\lp \omega_1a_{1\mu}+\omega_2a_{2\mu} \rp (n\cdot x)
\end{gather}

Substitution of all relevant relations into the expression for the transition probability, and integration over the $\phi_w,\, \lambda$ parameters, results in the final expression,

\begin{gather}\label{eq:Wconstfinal}
W_{\text{beam}} =-\mfrac{e^2}{2\gamma}\int_0^{\infty} \nths\mfrac{du}{(1+u)^2} \ls \int dz +\mfrac{1+(1+u)^2}{z\,(1+u)} \frac{d}{dz}\rs \Ai(z) \notag\\[4pt]
\text{where} \quad\quad z\equiv \ls\mfrac{u^2}{(\Upsilon_1 \vec{\hat a}_1+\Upsilon_2 \vec{\hat a}_2)^2}\rs^{1/3}
\end{gather}

In $W_{\text{beam}}$, the argument of the Airy function, $z$, is the only object that depends on the field strengths of the charge bunch fields. These field strengths are expressed by the parameters $\Upsilon_1,\,\Upsilon_2$, where $\vec{\hat a}_1,\vec{\hat a}_2$ are unit 3-vectors lying along each bunch field's 3-potential. \\

The combination $(\Upsilon_1 \vec{\hat a}_1+\Upsilon_2 \vec{\hat a}_2)^2$ is the Lorentz invariant quantity, $\Upsilon$, of equation \ref{eq:Upsilon2f}. $W_{\text{beam}}$ reduces to the usual beamstrahlung expression when the field of the initial particle's charge bunch is neglected ($\Upsilon_1\rightarrow 0$). \\

Since the transition rate is a Lorentz invariant quantity, the final analytic form of $W_{\text{beam}}$ is the same in all reference frames. The reference frame usually considered for collider processes, is the centre of mass frame of the colliding electron-positron bunches. \\

\begin{figure}[htb]
\centering
\includegraphics[width=0.45\textwidth]{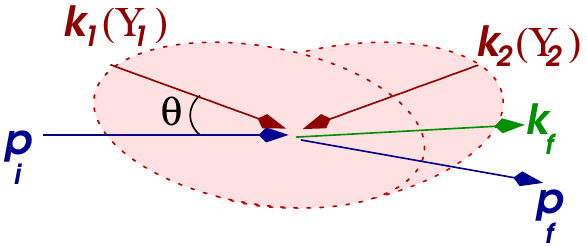}
\caption{\bf Beamstrahlung process in two overlapping charge bunch fields}\label{fig:beamstrpic}
\end{figure}

The $\Upsilon_1,\,\Upsilon_2$ parameters, one for each of the charge bunches (see figure \ref{fig:beamstrpic}), are effectively ratios of the field strength experienced by an individual charge, to the Schwinger critical field strength of $E_{\text{cr}}=1.32\times10^{18}$ V/m. \\

With $\gamma E_{1,2}$ being the relativistically boosted field strengths of the bunches in the centre of mass frame, $n_{1,2}$ being unit null 4-vectors representing the propagation direction of each bunch, and the unit 4-momentum of an individual charge $\hat{p}_i\equiv(1,\vec{p}_i/\epsilon_i)$, 

\begin{gather}
\Upsilon_{1,2}=\mfrac{e|\vec{a}_1| (k_1\cdot p_i)}{m^3}\equiv\mfrac{\gamma E_{1,2}}{E_{\text{cr}}}(n_{1,2}\cdot \hat{p}_i)
\end{gather}

Typically for planned future linear colliders such as the International Linear Collider \cite{ilctdr13}, $\gamma E_{\text{1,2}}/E_{\text{cr}}$ is of the order $0.1$. So $\gamma E_{\text{1,2}}/E_{\text{cr}}=0.1$ is used here in a numerical consideration of the beamstrahlung transition rate.\\

\begin{figure}[htb]
\centering
\includegraphics[width=0.5\textwidth]{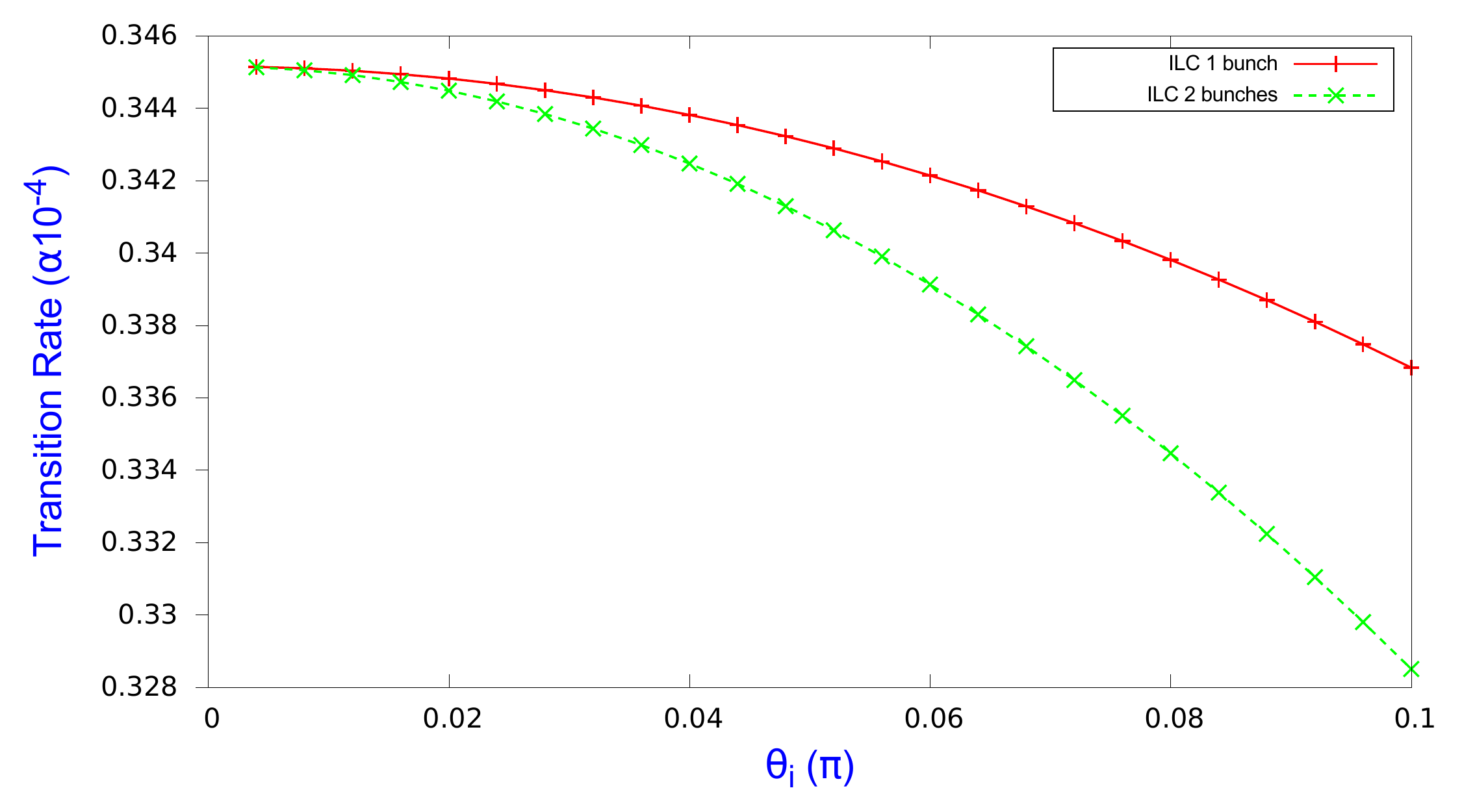}
\caption{\bf beamstrahlung transition rate per unit time due to initial particle angle for 250 GeV electrons and $\gamma E_{\text{1,2}}/E_{\text{cr}}=0.1$}\label{fig:2fpowspect}
\end{figure}

If allowance is made for an angular variation of an individual radiating particle with respect to the propagation direction of its own bunches ($\theta$ in figure \ref{fig:beamstrpic}), it is possible to produce a numerical comparison of the beamstrahlung due to the oncoming bunch only with the beamstrahlung that takes into account both charge bunch fields. Figure \ref{fig:2fpowspect} shows the beamstrahlung transition probability as the angle $\theta$ varies. \\

A full analysis of the significance of the new analytic form for the beamstrahlung transition rate is to be left to future work. Nonetheless the aim of this section, that of showing that fermion solutions in two overlapping charge bunch fields lead to analytic and numeric differences in transition rates, has been achieved. Moreover, the use of fermion solutions in two charge bunch fields within the Furry picture, provides a methodology for calculating the transition rate of other collider processes in the presence of the intense electromagnetic fields of colliding charge bunches.  

\section{Conclusions}\label{sect:concl}

The field strengths of the charge bunches of future lepton colliders are strong enough to potentially affect all collider processes taking place in their presence. This has prompted a desire to theoretically treat particle physics processes that occur in the overlapping bunch fields at the interaction point, non-perturbatively with respect to the charge bunch fields. Such a treatment is provided by the Furry picture, which calculates the exact contribution from the fields. The Furry picture reduces to the normal interaction picture in the limiting case of vanishing external field strength. \\

Employment of the Furry picture at collider interaction points, requires solutions of the field equations of charged particles coupled to both charge bunch fields. Since these fields are tied to massive sources, it is possible to choose a reference frame in which the charge bunches co-propagate and the problem simplifies. Solutions of the equations of motion for spin 1/2 particles in two co-propagating fields with co-planar 3-potentials were obtained. In the limiting case of orthogonal 3-potentials and the vanishing of one of the fields, the general solution reduces to previously obtained solutions \cite{Lyulka75,Pardy06,Volkov35}. \\

The fermion wavefunction in two co-propagating fields is written in terms of a null 4-vector whose 3-vector part has a unit norm and lies along the common propagation direction (equation \ref{eq:finalpsicp}). These solutions are orthogonal and complete on equal time hyperplanes. Canonical commutation relations and standard quantisation were assumed to follow from the orthogonality and completeness properties. \\

The solutions were used to calculate a more general form of the beamstrahlung transition rate for an electron radiating a photon in the fields of both colliding charge bunches. The analytic difference between the one charge bunch beamstrahlung and two charge bunch beamstrahlung is encapsulated by the Lorentz invariant parameter $\Upsilon$, which is a function of the electric field strengths of the charge bunch fields and the angle between them. \\

A numerical comparison of the beamstrahlung transition rate in one charge bunch field, to that in two charge bunch fields, showed differences for beam parameters expected at a future linear collider. This difference, depending on an angle between the propagation direction of the initial particle and that of its charge bunch, may be important for precision calculation required at linear colliders. A future detailed study will examine this process and its ramifications in greater detail.\\

A crucial question, given the precision physics program planned for future linear colliders, is, what is the effect of the strong charge bunch fields on collider process in general? Building upon the work done here, it will be natural to consider transition rates of general physics processes using solutions of the equations of motion in collinear fields within the Furry interaction picture. \\

Apart from the beamstrahlung, the coherent pair production, photon absorption and decays in final states can all be considered first order Furry picture processes. Second order processes involving an incoming electron and positron in the initial state, will necessarily interact with both charge bunch fields. The fermion solutions in two charge bunch fields used here, will be a prerequisite for the calculation of transition rates. \\

For ongoing work, the wavefunctions of charged spin 0 and spin 1 particles in two charge bunch fields will have to be specified. Two point Furry picture correlation functions leading to the particle propagators in two charge bunch fields are also required, as are techniques needed to calculate higher order Furry picture transition rates. This program of work will be embarked upon in forthcoming papers.

\acknowledgement{This work was supported in part by DFG Sonderforschungsbereich SFB 676}

\appendix
\renewcommand\thesection{Appendix \Alph{section}}
\section{Solution of the Dirac equation in two collinear fields}\label{sect:App1}

Required, is the wavefunction solution, $\psi_{\text{CP}}$, for a fermion embedded in two co-propagating, plane-wave, electromagnetic fields of general form. Linear combinations of the 4-potentials can be formed in order to achieve orthogonal components. With general functions $f_1,f_2$

\begin{align}\label{Areform}
A^e_\mu &= a_{1\mu} f_1(\phi_1)+a_{2\mu} f_2(\phi_2)\equiv A_1(\phi_1)+A_2(\phi_2) \notag\\
        &= \hat{a}_{1\mu}|\vec{a}_1| f_1(\phi_1)+\hat{a}_{2\mu} |\vec{a}_2| f_2(\phi_2) \notag\\
 & \equiv a_{+\mu} f_{+}(\phi_1,\phi_2)+a_{-\mu} f_{-}(\phi_1,\phi_2) \\
\text{where }  &  a_{\pm\mu}= \mfrac{\hat{a}_{1\mu}\pm \hat{a}_{2\mu}}{\sqrt{2}} \;\;,\;\; f_{\pm}=\mfrac{|\vec{a}_1| f_1\pm|\vec{a}_2| f_2}{\sqrt{2}} \notag \\
 a_{\pm}\cd k_{1,2}&=a_{+}\cd a_{-}=k_1\cd k_2=0,\quad \phi_{1,2}\equiv k_{1,2}\cd x \notag \\
 & \hat{a}_1^2=\hat{a}_2^2=1 \notag
\end{align}

If $\psi_{\text{CP}}$ is a solution of the Dirac equation minimally coupled to the external field, it is also a solution of the second order Dirac equation,

\begin{gather}\label{eq2nddirac}
\ls\lp i\st{\partial}-eA^e\rp^2+m^2 + \mfrac{ie}{2}F^{\mu\nu}\sigma_{\mu\nu} \rs \psi_\text{CP}=0 \notag\\
\text{with}\quad F^{\mu\nu}=\partial^\mu A^{e\nu}-\partial^\nu A^{e\mu},\quad 
\sigma_{\mu\nu}=\mfrac{1}{2}(\gamma_\mu\gamma_\nu-\gamma_\nu\gamma_\mu)  \notag 
\end{gather}

The reformed external field 4-potential and the orthogonal relations of equation \ref{Areform} allow separable equations to be formed,

\begin{gather}
\ls\lp i\slashed{\partial}-ea_{\pm}f_{\pm}\rp^2+m^2 + ie\st{\partial}f_{\pm}\st{a}_{\pm} \rs \psi_\text{CP}=0 
\end{gather}

There are two sets of solutions corresponding to positive and negative energy solutions. Proposing $\psi^\pm_{\text{\tiny CP}}$ as a product of the free, fermion, plane wave solutions and general functions of the external field, $G_{\pm}\lp \phi_1,\phi_2\rp$, 

\begin{gather}\label{eq:Aplussol}
 \psi^\pm_\text{CP}(\phi_1,\phi_2)= u^\pm_{rp}\,e^{\mp i\ls p\cdot x + G_{+}\lp \phi_1,\phi_2\rp + G_{-}\lp \phi_1,\phi_2\rp\rs} 
\end{gather}

substitution into the second order Dirac equation yields first order partial differential equations,

\begin{gather}\label{eqpde}
\ls 2\lp k_1\cd p\rp \mfrac{d}{d\phi_1}+2\lp k_2\cd p\rp \mfrac{d}{d\phi_2}\rs G_\pm(\phi_1,\phi_2)= F_\pm(\phi_1,\phi_2) \notag\\
 \text{where}\; F_\pm\equiv 2e\lp a_{\pm}\cd p\rp f_{\pm} - e^2a_\pm^2 f_\pm^2+ie\st{\partial}f_{\pm}\st{a}_{\pm}
\end{gather}

Choosing a structure for the unknown functions $G^{\pm}$ that matches that of the known functions $F^{\pm}$, separate differential equations can be formed,

\begin{gather}\label{eq:splitde}
G_{\pm}(\phi_1,\phi_2)\equiv g_{\pm1}(\phi_1)+g_{\pm2}(\phi_2)+g_{\pm3}(\phi_1,\phi_2) \notag\\[4pt]
 \mfrac{dg_{\pm1}}{d\phi_1}=\mfrac{2e\lp a_{\pm}\cd p\rp |\vec{a}_1|f_{1} - e^2a_\pm^2 |\vec{a}_1|^2f_{1}^2+ief_1|\vec{a}_1|\st{k}_1\st{a}_\pm}{2\sqrt{2}\lp k_1\cd p\rp} \notag\\[4pt]
 \mfrac{dg_{\pm2}}{d\phi_2}=\mfrac{\pm2e\lp a_{\pm}\cd p\rp |\vec{a}_2|f_{2} - e^2a_\pm^2 |\vec{a}_2|^2f_{2}^2 \pm ief_2|\vec{a}_2|\st{k}_2\st{a}_\pm}{2\sqrt{2}\lp k_2\cd p\rp} \\[4pt]
\ls 2\lp k_1\cd p\rp \mfrac{d}{d\phi_1}+2\lp k_2\cd p\rp \mfrac{d}{d\phi_2}\rs g_{\pm3}= \mp \mfrac{e^2a_\pm^2|\vec{a}_1||\vec{a}_2|}{\sqrt{2}}f_1\,f_2 \notag
\end{gather}

These differential equations are easily solved. Forming the sum $G_{+}+G_{-}$, the positive and negative Dirac wavefunction solutions in two co-propagating plane wave electromagnetic fields of general form are,

\begin{gather}\label{eq:psiCP}
\psi^{\pm}_{\text{CP}}(\phi_1,\phi_2) = cE^\pm_{1p}(\phi_1)E^\pm_{2p}(\phi_2)\; u^\pm_{rp}
e^{\mp i \ls p\cdot x - e^2 a_1\cdot a_2H \rs } \\
E_{ip}^\pm(\phi_i)\equiv\ls 1 \mp \mfrac{e\st{A}_i^e\st{k}}{2(k_i\cd p)}\rs e^{\mp i\mathlarger{\int}^{\phi_i}\mfrac{2eA_i(\xi_i)\cdot p \mp e^2A_i(\xi_i)^2}{2k_i\cdot p}d\xi_i} \notag \\
H\equiv \int^{\phi_1}\!\mfrac{d\chi}{k_1\cd p}\, f_1[\chi]\;f_2\!\ls \mfrac{k_2\cd p}{k_1\cd p}\lp \chi \!-\! \phi_1\rp +\phi_2\rs \notag
\end{gather}

Limiting cases of the above general solution were obtained earlier in the literature. Restricting the 3-potentials to orthogonality $a_1\cdot a_2=0$, the term containing the $H$ function drops out and a product of Volkov solutions is obtained \cite{Lyulka75,Pardy06}. Additionally, a single Volkov solution results from removing one of the fields by setting $|\vec{a}_1|$ or $|\vec{a}_2|$ to zero. \\

For collinear fields, the energy can be extracted from photon 4-momenta leaving a null 4-vector $n_\mu=(1,\vec{n})$, whose 3-part is parallel to the collinear direction

\begin{gather}
k_{1,2\;\mu}\!=\!\omega_{1,2}\;n_\mu,\;\; \phi_{1,2}\!=\!\omega_{1,2}\;n\cd x,\;\;k_{1,2}\cd p\!=\!\omega_{1,2}\;n\cd p
\end{gather}

In equation \ref{eq:psiCP}, making shifts of integration variables, $\xi_i\!\rightarrow\!\omega_i\xi_i,\;\;\chi\!\rightarrow\!\omega_1\chi$ and gathering terms, the general solutions reduce to a Volkov solution in terms of the 4-vector $n_\mu$,

\begin{gather}
\psi^{\pm}_{\text{CP}}(n\cd x) = c\; J^\pm_{px}\; u^\pm_{rp}\, e^{\mp i\,p\cdot x} \\
J^\pm_{px}\equiv\ls 1 \mp\mfrac{\st{A}^e\!(n\cd x)\st{n}}{n\cd p}\rs\, e^{\mp i\mathlarger\int^{n\cdot x}\small\mfrac{2A^e(\xi)\cd p \mp A^e(\xi)^2}{2n\cd p}d\xi} \notag
\end{gather}

\bibliographystyle{unsrt}
\bibliography{/home/hartin/Physics_Research/mypapers/hartin_bibliography}

\begin{thebibliography}{10}

\bibitem{Baier68}
V.M. Baier, V.N.~Katkov and V.M. Strakhovenko.
\newblock {\em Sov Phys JETP}, 26:854, 1968.

\bibitem{Baier69}
V.M. Baier, V.N.~Katkov and V.M. Strakhovenko.
\newblock {\em Sov Phys JETP}, 28:807, 1969.

\bibitem{YokChe91}
K.~Yokoya and P.~Chen.
\newblock Beam-beam phenomena in linear colliders.
\newblock Technical report, KEK Preprint 91-2, 1991.

\bibitem{Furry51}
W.H. Furry.
\newblock {\em Phys Rev}, 81:115, 1951.

\bibitem{Ritus72}
V.I. Ritus.
\newblock Radiative corrections in quantum electrodynamics with intense field
  andtheir analytic properties.
\newblock {\em Ann Phys}, 69:555--582, 1972.

\bibitem{Hartin09}
A.~Hartin.
\newblock On the equivalence of semi-classical methods for {QED} in intense
  external fields.
\newblock {\em J Phys Conf Ser}, 198:012004, 2009.

\bibitem{AlsBau97}
K.~Trautmann~D. Alscher, A.~Hencken and G.~Baur.
\newblock Multiple electromagnetic electron positron pair production in
  relativistic heavy ion collisions.
\newblock {\em Phys Rev A}, 55:396, 1997.

\bibitem{Budnev75}
I.F. Budnev, V.M.~Ginzburg and et~al.
\newblock The two-photon particle production mechanism. physical problems.
  applications. equivalent photon approximation.
\newblock {\em Physics Reports}, 15:181--282, 1975.

\bibitem{BagGit90}
V.G. Bagrov and D.M. Gitman.
\newblock {\em Exact solutions of relativistic wave equations}.
\newblock Kluver Academic Publishers, 1990.

\bibitem{Kurilin99}
A.V. Kurilin.
\newblock Particle physics in intense electromagnetic fields.
\newblock {\em Nuovo Cim A}, 112A:977--1000, 1999.

\bibitem{Lyulka75}
V.A. Lyul'ka.
\newblock {\em Sov Phys JETP}, 40(5):815, 1975.

\bibitem{Pardy06}
M~Pardy.
\newblock {\em Int J Theor Phys}, 45:647--659, 2006.

\bibitem{NarFof96}
N.B. Narozhnyi and M.S. Fofanov.
\newblock Photon emission by an electron in a collision with a short focused
  laser pulse.
\newblock {\em JETP}, 83(1):14--23, 1996.

\bibitem{BerVal73}
I.~Berson and J.~Valdmanis.
\newblock Electron in the field of two monochromatic electromagnetic waves.
\newblock {\em J Math Phys}, 14:1481, 1973.

\bibitem{SenGupta67}
N.D. Sen~Gupta.
\newblock {\em Z Phys}, 200:13, 1967.

\bibitem{Jackson75}
J.D. Jackson.
\newblock {\em Classical Electrodynamics}.
\newblock John Wiley and Sons, New York, 1975.

\bibitem{Schweber62}
S.~Schweber.
\newblock {\em An introduction to Relativistic Quantum Field Theory}.
\newblock Harper and Row, 1962.

\bibitem{JauRoh76}
J.M. Jauch and F.~Rohrlich.
\newblock {\em The Theory of Photons and Electrons}.
\newblock Springer-Verlag, Berlin, 1976.

\bibitem{Moort09}
G.~Moortgat-Pick.
\newblock The {F}urry picture.
\newblock {\em J Phys Conf Ser}, 198:012002, 2009.

\bibitem{FraGitShv91}
D.M. Fradkin, E.S.~Gitman and M.~Shvartsman.
\newblock {\em Quantum Electrodynamics with unstable vacuum}.
\newblock Springer-Verlag, 1991.

\bibitem{Volkov35}
D.M. Volkov.
\newblock {\"U}ber eine {K}lasse von {L}{\"o}sungen der {D}iracschen
  {G}leichung.
\newblock {\em Zeitschrift f{\"u}r Physik}, 94:250--260, 1935.

\bibitem{Ritus79}
V.~Ritus.
\newblock Quantum effects of the interaction of elementary particles with an
  intense electromagnetic field.
\newblock {\em Journal of Soviet Laser Research}, 6 (5):497--617, 1985.

\bibitem{Filip85}
P.~Filipowicz.
\newblock Relativistic electron in a quantised plane wave.
\newblock {\em J Phys A}, 18:1675--1685, 1985.

\bibitem{BocFlo10}
M.~Boca and V.~Florescu.
\newblock The completeness of {V}olkov spinors.
\newblock {\em Rom. Phys J}, 55:511--525, 2010.

\bibitem{Hartin11a}
A.~Hartin and G.~Moortgat-Pick.
\newblock High intensity {C}ompton scattering in a strong plane-wave field of
  general form.
\newblock {\em EPJ C}, 71:1729, 2011.

\bibitem{ilctdr13}
T.~Behnke~et al.
\newblock International linear collider technical design report.
\newblock Technical report, ILC Global Design Effort, 2013.

\end{thebibliography}

\end{document}